\documentclass[12pt]{iopart}

\usepackage{graphicx}

\begin{document}

\title{A renormalized Gross-Pitaevskii Theory and vortices in a
strongly interacting Bose gas}

\author{Ch. Moseley}
\ead{Christopher.Moseley@Physik.Uni-Augsburg.de}
\address{Institut f\"ur Physik, Universit\"at Ausgburg, D-86135 Augsburg, Germany}

\author{K. Ziegler}
\address{Institut f\"ur Physik, Universit\"at Ausgburg, D-86135 Augsburg, Germany}

\date{\today}
\begin{abstract}
We consider a strongly interacting Bose-Einstein condensate in a spherical
harmonic
trap. The system is treated by applying a slave-boson representation for
hard-core
bosons. A renormalized
Gross-Pitaevskii theory is derived for the condensate wave function that 
describes
the dilute regime (like the conventional Gross-Pitaevskii theory) as well as the
dense regime. We calculate the condensate density of a rotating condensate 
for both the
vortex-free condensate and the condensate with a single vortex
and determine the critical angular velocity for the formation of a stable 
vortex in a rotating trap.
\end{abstract}


\maketitle

\section{Introduction}

In this paper we shall study two aspects of a strongly interacting Bose gas at
high density. One is related to a consistent treatment of a strongly
interacting Bose gas in terms of
an effective Gross-Pitaevskii (GP) Eq. with renormalized parameters. The second
aspect is related to the formation of a vortex in a trapped condensate in the presence
of strong interaction. 

The stationary form of the conventional GP Eq. 
\begin{equation}
 \left[-\frac{\hbar^2}{2m} \nabla^2 - \mu + V({\bf r}) +
 g|\Phi({\bf r})|^2 \right] \Phi({\bf r}) = 0 \;
 \label{conventionalGP}
\end{equation}
describes the condensate order parameter $\Phi$ of a Bose gas in a trapping
potential at zero temperature,
where $\mu$ is the chemical potential and $g$ the repulsive coupling constant
\cite{Dalfovo99,pitaevskii,leggett}. In the absence of a trapping potential, a solution 
of Eq. (\ref{conventionalGP}) is given by
\begin{equation}
 |\Phi|^2 = \frac\mu g .
 \label{gpe}
\end{equation}
This describes a linearly increasing condensate density (the latter is proportional
to $|\Phi|^2$) with respect to the chemical potential.
Although it takes the repulsion into account
by a factor $1/g$ which is decreasing  with increasing coupling constant $g$, the
saturation of $n_0$
cannot be seen in this solution. From the physical point of view, in a realistic
description for large densities, the particle density must saturate because there is
a finite scattering volume around each particle. Furthermore, for increasing particle
density, the condensate density should reach a maximum and for even larger densities,
decrease again until its total destruction, because of the increasing interparticle
interaction. This behaviour has also been found by variational perturbation theory 
\cite{kleinert2} and diffusion Monte Carlo calculations \cite{dubois1}.
In other words, the strong effect of the repulsion
in a dense condensate is not really described by the conventional GP Eq.
In order to describe condensates at higher densities, the second order term in the
low-density expansion of the energy density has been taken into account which lead to a
modified GP theory \cite{Dalfovo99,dubois1,fabrocini1,nilsen1}.

Although in many experimentally realized situations the BEC is in the weakly interacting
regime where it is well described by GP theory, it might be possible to reach the
strongly interacting regime. The main problem at high particle densities is
the instability of the Bose gas by the formation of molecules due to three-particle
interactions \cite{gammal3}. Here, we will assume that molecule formation does not occur. 
This might be unrealistic for some systems, but in others it is not, e.g. for Bose gases 
in optical lattices.

It has been shown that the slave-boson approach to a
hard-core Bose gas provides a
mean-field Eq., similar to the GP Eq., that leads to a saturation of the condensate
$n_0\le1$ \cite{kotliar86,ziegler1,ziegler2,Stoof,Lu}. Here we will discuss a simplified
version of this mean-field
equation which is the same type of nonlinear Schr\"odinger Eq. like in the GP
approach. However, in contrast to the latter the parameters are renormalized such that
Eq. (\ref{gpe}) becomes
\[
|\Phi_{\rm R}|^2=\frac{\mu_{\rm R}}{g_{\rm R}},
\]
where the renormalized parameters $\mu_{\rm R}$ and $g_{\rm R}$ are functions of the bare
parameters $\mu$ and $n_0$. At the phase transition between the non-condensed Bose gas
and the BEC $\mu_{\rm R}$ vanishes, then increases linearly with increasing $\mu$, reaches a
maximum and decreases again until the condensate is destroyed totally due to strong
interaction effects. We use this approach to calculate the condensate density profile
of a trapped BEC.

The formation of vortices in a rotating condensate is understood as a local destruction
of the condensate. From the depletion effect due to strong interaction it can be
anticipated that the tendency of vortex formation is enhanced in a dense Bose gas.
This implies a reduction of the critical angular velocity $\Omega_c$ with increasing
density by the interaction. Vortex formation in rotating traps in the region beyond
the validity of GP has also been studied by the modified GP theory and by
variational Monte Carlo methods \cite{nilsen1}.

The paper is organized as follows: In the following section we survey the results of the
slave-boson approach for a hard-core Bose gas on a lattice and derive a corresponding
mean field theory for a continuous system in an external trap potential. 
Then we derive the renormalized GP Eq.
and evaluate solutions with and without a straight vortex and determine the critical
angular velocity for the vortex formation.

\section{The Model}

The slave-boson representation of a strongly interacting Bose gas 
is based on the idea that the bosons fill the space with finite
density, where each particle occupies a lattice cell related to a sphere
with radius $a_s$ ($s$-wave scattering length) \cite{ziegler1}. 
Singly occupied and empty sites are
described by two complex fields $b_x$ and $e_x$, respectively \cite{ziegler2}.
Here, $b^\ast_x$ and $b_x$ represent a creation and annihilation process of a boson at site $x$,
while $e^\ast_x$ and $e_x$ represent the creation and annililation of a ``hole''.
In the functional integral representation \cite{popov}, the grand canonical 
partition function in classical approximation is given as
\begin{displaymath}
 \fl Z = \int \exp\left[-\beta \sum_{x,x'}b_x^\ast e_x t_{x,x'} e_{x'}^\ast b_{x'}
 + \beta \sum_x \mu_x|b_x|^2\right] 
\end{displaymath}
\begin{equation}
 \times
\prod_x \delta(|e_x|^2+|b_x|^2-1) {\rm d}b_x {\rm d}b_x^\ast {\rm d}e_x 
{\rm  d}e_x^\ast \; ,
 \label{Z}
\end{equation}
where $1/\beta=k_{\rm B}T$ is the thermal energy.
In a $d$-dimensional lattice,
\begin{displaymath}
 \hat{t}_{x,x'} = \left\{ \begin{array}{l@{\quad}l}
 -J/(2d) & \mbox{if $x,x'$ nearest neighbours}\\
 0 & \mbox{else} \end{array} \right.
\end{displaymath} 
is the nearest-neighbour tunneling rate and $\mu_x=\mu-V_x$ is the
space-dependent
chemical potential that includes the trapping potential $V_x$. Instead of $\mu_x$ we
will write only $\mu$ subsequently and assume implicitly that the effective chemical
potential can depend on space. 
The fields $e$ and $b$ are dimensionless. Moreover, we rescale all physical
energies by a multiple of the hopping rate $\alpha J$ to obtain dimensionless 
quantities:
\begin{equation}
{\hat t}\to \hat{t}' = \frac{1}{\alpha J}\hat{t}\; , 
\quad \mu\to \mu' = \frac{1}{\alpha J} \, \mu \; ,
\quad \beta\to\beta' =\alpha J \beta \; .
\end{equation}
The hopping term is quartic in the field
variables, due to the fact that a hopping process is characterized as an ``exchange process''
of a boson and an empty site. 
The $\delta$ function enforces the constraint that each lattice site is either singly
occupied or empty but excludes a multiple occupation. 

A similar approach has been applied to the Bose-Hubbard model which allows multiple
occupation, by introducing additional fields, one for each occupation number 
\cite{Stoof,Lu}. The well-known zero-temperature phase diagram with Mott-insulating 
phases for integer lattice fillings $n=0,1,2,\ldots$ (``lobes'') was found. 
In contrary, our hard-core Boson model is restricted to the two lattice fillings $n=0,1$ which
is a simplification, but contains all relevant aspects of Bose-Einstein condensation with
repulsive interaction.
A second simplification of our approach is, that we neglect quantum fluctuations
by treating the grand-canonical partition function in classical approximation. This allows
us to integrate out the constraint exactly. On the other hand, it restricts the
applicability of our approach to non-zero temperatures.

Two new fields are introduced by a Hubbard-Stratonovich transformation,
the complex field $\Phi_x$ which describes the condensate wave function, and the real field
$\varphi_x$ which is related to the total density of bosons \cite{ziegler1}:
\begin{displaymath}
 \fl Z = \int \exp \Bigg\{ - \beta' \Bigg[
 \sum_{x,x'} \Phi_{x}^\ast \left({\bf 1}-\hat{t}'\right)_{x,x'}^{-1}
 \Phi_{x'} + \sum_{x} \varphi_{x}^2 
\end{displaymath}
\begin{displaymath}
 + \sum_{x} (e_{x},b_{x}) \left(
 \begin{array}{cc} 2\varphi_{x} + 1 & \Phi_{x} \\ \Phi_{x}^\ast & -\mu'
 \end{array}\right)\left( \begin{array}{c}
  e_{x}^\ast \\ b_{x}^\ast \end{array}\right) \Bigg] \Bigg\}
\end{displaymath}
\begin{equation}
  \times
\prod_x \delta(|b_x|^2+|e_x|^2-1) {\rm d}b_x {\rm d}b_x^\ast
 {\rm d}e_x {\rm d}e_x^\ast {\rm d}\Phi_x {\rm d}\Phi_x^\ast {\rm d}\varphi_x \; .
 \label{ZHST}
\end{equation}
Integration over the fields $\Phi$ and $\varphi$ leads back to Eq. (\ref{Z}).
On the other hand, the fields $b_x$ and $e_x$ can be integrated out in
(\ref{ZHST}) because they appear in the exponent as quadratic forms. This leads to the 
partition function
\begin{equation}
 Z=\int e^{-S_b-S_1} \prod_x {\rm d}\Phi_x {\rm d}\Phi_x^\ast
\label{part01}
\end{equation}
with the kinetic part of the action
\begin{equation}
 S_b=\beta'\sum_{x,x'} \Phi_x \left({\bf 1}-\hat{t}'\right)_{x,x'}^{-1} 
 \Phi_{x'}^\ast
 \label{Sdiscrete_kin}
\end{equation}
and the potential part
\begin{equation}
 \fl S_1=-\sum_x \log \left( \int_{-\infty}^\infty e^{-\beta' \varphi_{x}^2} \,
 \frac{ \sinh\left[\beta'\sqrt{\left(\varphi_{x} +\frac{\mu'}2 \right)^2 
 + |\Phi_{x}|^2}\right]}{ \beta'\sqrt{\left(\varphi_{x}
 + \frac{\mu'}2\right)^2 + |\Phi_{x}|^2}} {\rm d}\varphi_{x} \right) := -\sum_x Z_x \; .
 \label{Sdiscrete_pot}
\end{equation}
The condensate density can be identified with
\begin{equation}
 n_0 = \frac{|\Phi_x|^2}{(1+1/\alpha)^2} \; ,
 \label{n0-ident}
\end{equation}
like argued in Appendix A, and the total particle density is given by the expectation value
\cite{ziegler1}
\begin{equation}
 n_{\rm tot} = \langle \varphi_x \rangle + \frac 12 \; .
 \label{total_density}
\end{equation}
We apply a saddle-point approximation to the integration in 
Eq. (\ref{part01}). This is controlled by the minimized action,
which means that we have to solve the equation
\begin{displaymath}
 \frac{\partial S}{\partial \Phi^\ast_x} =0 \; .
\end{displaymath}
This yields the mean-field equation
\begin{equation}
 \beta' \sum_{x'} \left({\bf 1}-\hat{t}'\right)_{x,x'}^{-1} \Phi_{x'}
 - \left[ \frac\partial{\partial(|\Phi_x|^2)} \log Z_x \right] 
 \Phi_x = 0 \; .
 \label{MF_discrete}
\end{equation}

In order to derive a mean-field equation which is applicable to a continuous
trapping potential, we perform the continuum approximation of (\ref{MF_discrete}).
If the field $\Phi_{\bf r}$ is varying only very slowly between neighbouring 
lattice sites, we can approximate
\begin{displaymath}
 \left({\bf 1}-\hat{t}'\right)_{x,x'}^{-1} \approx \frac 1{1+1/\alpha} \left(
 \delta_{x,x'} + \frac{1}{1+1/\alpha} \, \alpha J 
 \left(J \,\delta_{x,x'} + \hat{t}_{x,x'}\right) \right) \; ,
\end{displaymath}
and perform the substitution
\begin{displaymath}
 \sum_{x'} (J\,\delta_{x,x'}+\hat{t}_{x,x'}) \rightarrow -J a^2 \nabla^2 \; .
\end{displaymath}
This leads to the equation
\begin{equation}
 \left[ - \frac{Ja^2}{6} \nabla^2 + (1+\alpha)J - \frac{(1+1/\alpha)^2}{\beta} \,
 \frac\partial{\partial(|\Phi({\bf r})|^2)} \log Z({\bf r}) \right] \Phi({\bf r}) = 0 \; .
 \label{SBoriginal}
\end{equation}
with
\begin{displaymath}
 Z({\bf r}) = \int e^{-\beta' \varphi_{\bf r}^2} \, \frac{
 \sinh \left[\beta'\sqrt{(\varphi_{\bf r} + \mu'/2)^2+|\Phi({\bf r})|^2}\right]}{\left[ \beta'
 \sqrt{(\varphi_{\bf r} + \mu'/2)^2+|\Phi({\bf r})|^2}\right]} \, {\rm d}\varphi_{\bf r}
 \label{function_Z}
\end{displaymath}
for the spacially dependent order parameter $\Phi({\bf r})$
in a three-dimensional space and $a$ is the lattice constant of the discrete system.
In the continuum $a$ looses its identity as lattice constant, but describes a
characteristic length scale in Eq. (\ref{SBoriginal}), and can be interpreted as
the spacial extension of a boson. Thus, it should be of the same order of magnitude as
the $s$-wave scattering length $a_s$ \cite{Dalfovo99}. Eq. (\ref{SBoriginal})
is the analogue of the GP Eq. in the case of our
slave-boson approach. The parameters can be identified with those of the
conventional GP Eq.: The mass $m$ of the particles is given by the hopping constant 
$J$ and the lattice constant $a$ via
\begin{equation}
 \frac{\hbar^2}{2m}\equiv\frac{Ja^2}6 \; .
 \label{mass}
\end{equation}

\section{Renormalized Gross-Pitaevskii Equation}

\begin{figure}
\centering
\includegraphics{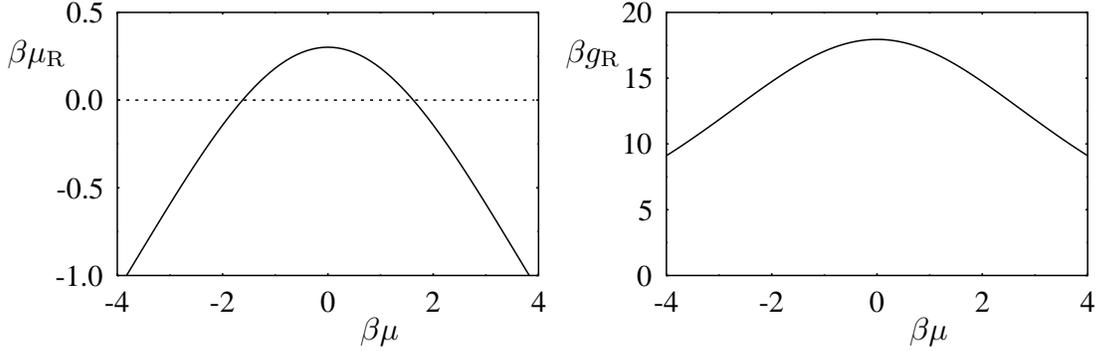}
\caption{Coefficients $\mu_{\rm R}$ and $g_{\rm R}$ of the
RGP theory plotted against the chemical potential $\mu$ for the tunneling rate $\alpha J=k_{\rm B}T$.
Both functions are symmetric in $\mu$. A BEC exists if $\mu_{\rm R}>0$. The two points
where $\mu_{\rm R}=0$ mark the phase transition between the non-condensate and the BEC.
The coefficient $g_{\rm R}$ ist always positive.}
\label{coeff}
\end{figure}

The continuum limit of the action defined by Eq. (\ref{Sdiscrete_kin}) and
(\ref{Sdiscrete_pot}) is
\begin{equation}
 S = \int \left\{ \beta' \Phi^\ast({\bf r}) \left[
 -\frac{\alpha}{(1+\alpha)^2} \frac{a^2}{6} \nabla^2 + \frac{1}{1+1/\alpha}
 \right] \Phi({\bf r}) - \log Z({\bf r}) \right\} \, {\rm d}^d r \; .
 \label{Scontinuous}
\end{equation}
Applying the variational principle
\begin{displaymath}
 \frac{\partial S}{\partial \Phi^\ast_x} =0 \; ,
\end{displaymath}
we obtain the full mean-field Eq. (\ref{SBoriginal}) directly.
If the order parameter $\Phi$ is small, we can expand the potential part of 
the action up to fourth order:
\begin{equation}
 \frac{1}{1+1/\alpha}|\Phi|^2- \frac 1{\beta'} \log Z({\bf r}) = a_0(\mu') + a_2(\mu') |\Phi|^2 +
 \frac 12 a_4(\mu') |\Phi|^4 + {\cal O}(|\Phi|^6) \; ,
 \label{RGP}
\end{equation}
where we have introduced the coefficients
\begin{eqnarray}
 a_0(\mu') &=& - \left. \frac 1{\beta'} \log Z({\bf r}) \right|_{\Phi=0} \label{coeff_a0}\\
 a_2(\mu') &=& - \left. \frac 1{\beta'} \frac\partial{\partial|\Phi|^2} 
 \log Z({\bf r}) \right|_{\Phi=0} + \frac{1}{1+1/\alpha} \label{coeff_a2}\\
 a_4(\mu') &=& - \left. \frac 1{\beta'} \frac{\partial^2}{(\partial|\Phi|^2)^2} 
 \log Z({\bf r}) \right|_{\Phi=0} \; .
 \label{coeff_a4}
\end{eqnarray}
Further, we introduce the rescaled field
\begin{displaymath}
 \Phi_{\rm R}({\bf r}) = \frac 1{1+1/\alpha} \, \Phi({\bf r}) \; ,
\end{displaymath}
which we identify with the condensate wave function of our renormalized
GP theory. We now introduce the renormalized coefficients
\begin{displaymath}
 \mu_{\rm R}(\mu,J) \equiv - \frac{(1+\alpha)^2}{\alpha}J a_2(\mu')
\end{displaymath}
and
\begin{displaymath}
 g_{\rm R}(\mu,J) \equiv \frac{(1+\alpha)^4}{\alpha^3} J a_4(\mu') \; .
\end{displaymath}
After neglecting the term of order $|\Phi|^6$ in the expansion (\ref{RGP}), we get
the renormalized Gross-Pitaevskii (RGP) Eq.
\begin{equation}
 \left[-\frac{Ja^2}{6}\nabla^2 - \mu_{\rm R}(\mu,J) +
 g_{\rm R}(\mu,J)|\Phi_{\rm R}({\bf r})|^2 \right] \Phi_{\rm R}({\bf r}) = 0 \; . 
 \label{rgpe}
\end{equation}
It has the same form as the conventional GP Eq. (\ref{conventionalGP}), when
$\mu$ and $g$ are replaced by $\mu_{R}$ and $g_{R}$.
Their $\mu$-dependence is plotted in Fig. \ref{coeff}.
In the case of a trapping potential, where the chemical potential $\mu$ is space-dependent,
$\mu_{\rm R}$ and $g_{\rm R}$ are space-dependent aswell.
While $g_{\rm R}$ is always positive, $\mu_{\rm R}$ can change sign. A BEC exists if
$\mu_{\rm R}>0$. The phase transition between the BEC and the non-condensate phase, 
i.e. the point at which the condensate order parameter vanishes, is given by the relation 
$\mu_{\rm R}(\mu,J)=0$.


\section{Results}


\subsection{Zero-temperature result}

\begin{figure}
\centering
\includegraphics{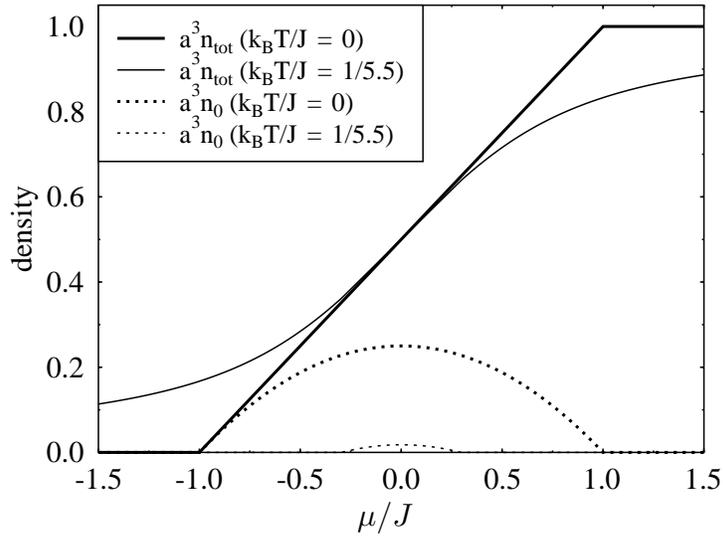}
\caption{Condensate density $n_0$ and total density $n_{\rm tot}$ of a translational
invariant system, plotted against chemical
potential for zero-temperature (thick lines) and finite temperature (thin lines). }
\label{density}
\end{figure}

In the zero-temperature limit we can integrate out the $\varphi$-field in Eq.
(8) exactly by a saddle-point integration, as shall be
shown in this paragraph. Therefore, we write
\begin{displaymath}
 Z({\bf r})=\frac 1{2\beta'} (Z_- - Z_+) \; ,
\end{displaymath}
where
\begin{displaymath}
 Z_\pm = \int_{-\infty}^\infty \frac{
 e^{-\beta' f_\pm(\varphi,|\Phi|^2)} }{ 
 \sqrt{(\varphi+\frac{\mu'}{2})^2+|\Phi|^2} } \, {\rm d}\varphi
\end{displaymath}
and
\begin{displaymath}
 f_\pm(\varphi,|\Phi|^2) = \varphi^2 \pm \sqrt{\left(\varphi+\frac{\mu'}{2}\right)^2+|\Phi|^2} \; .
\end{displaymath}
It is possible to perform a saddle-point approximation by expanding the functions $f_\pm$
up to second order in $\varphi$ around their minimum $\varphi_0$:
\begin{displaymath}
 f_\pm(\varphi,|\Phi|^2) = f_\pm(\varphi_0,|\Phi|^2) + \frac 12
 \frac{\partial^2 f_\pm}{\partial\varphi^2}(\varphi_0,|\Phi|^2)(\varphi-\varphi_0)^2 +
 {\cal O}(\varphi^2) \; .
\end{displaymath}
In the limit of large $\beta$, the saddle-point integration becomes exact and yields
\begin{displaymath}
 Z_\pm = \sqrt{\frac\pi{(\varphi_0+\frac{\mu'}{2})^2+|\Phi|^2}} \, 
 \frac{e^{-\beta' f_\pm(\varphi_0,|\Phi|^2)}}{
 \sqrt{ \frac{\beta'}2 \frac{\partial^2 f_\pm(\varphi_0,|\Phi|^2)}{\partial\varphi^2} }} \; .
\end{displaymath}
The minimum is found to satisfy the equation
\begin{displaymath}
 |\Phi|^2 = \left(\varphi_0+\frac{\mu'}{2}\right)^2 \left( \frac 1{4\varphi_0^2} -1 \right) \; ,
\end{displaymath}
and we have
\begin{displaymath}
 f_\pm(\varphi_0) = \varphi_0^2 - \frac 12 - \frac{\mu'}{4\varphi_0} \; ; \;
 \frac{\partial^2 f_\pm(\varphi_0)}{\partial\varphi^2} = 2 - 
 \frac{8|\Phi|^2 \varphi_0^3}{(\varphi_0+\frac{\mu'}{2})^3} \; .
\end{displaymath}
We find the following zero-temperature result for a translational invariant condensate
($\Phi_x\equiv\Phi={\rm const}$) from
the mean-field Eq. (\ref{SBoriginal}): For the condensate density we find
\begin{equation}
 n_0 = \frac{|\Phi|^2}{(1+1/\alpha)^2} = \left\{ \begin{array}{l@{\quad}l}
 \frac 14 \left(1-\frac{\mu^2}{J^2}\right) & \mbox{if } -J<\mu<J \\
 0 & \mbox{else} \; , \end{array} \right.
\end{equation}
and the total particle density given by Eq. (\ref{total_density}) is 
\begin{equation}
 n_{\rm tot} = \varphi_0+\frac 12 = \left\{ \begin{array}{l@{\quad\mbox{if }}l}
 0 & \mu\le -J \\
 \frac 12 \left(1-\frac\mu J\right) & -J<\mu<J \\
 1 & J\le \mu \; . \end{array} \right.
\end{equation}
The solution of the mean-field equation is plotted in Fig. \ref{density} for zero-temperature
and near the critical temperature $T_c$ where the BEC breaks down.
In the dilute regime at $T=0$, the chemical potential can be written as
$\mu=-J+\Delta\mu$, where $\Delta\mu\ll J$. In this limiting case we find
$\mu_{\rm R}=\Delta\mu+{\cal O}(\Delta\mu^2)$ and $g_{\rm R}=2J$, which is consistent with
the conventional GP Eq. with a shifted chemical potential.

\subsection{Vortex-free trapped condensate}

Assuming a dense condensate, where the repulsive interaction between bosons dominates
their kinetic energy, we neglect the differential term in Eqs. (\ref{SBoriginal}) and 
(\ref{Scontinuous}). This is called the Thomas-Fermi (TF) approximation \cite{Dalfovo99}.
In the following we use a spherical trapping potential
\begin{equation}
 V({\bf r})=\frac m2 \omega_{\rm ho}^2 {\bf r}^2 \; .
 \label{trap}
\end{equation}
In typical experiments, the oscillator length $d_{\rm ho}=\sqrt{\hbar/m\omega_{\rm ho}}$ is of the
order of a few $\rm \mu m$ \cite{Dalfovo99}, where $\omega_{\rm ho}$ is the trap frequency
measured in Hz. Considering, for instance, $^{85}$Rb atoms near a Feshbach resonance 
\cite{Cornish00}, we can study a Bose gas in a dense regime with a scattering length 
$a_s\sim a\sim 200\rm nm$. In our calculations we choose the parameters
\begin{equation}
 \beta'=1 \; , \quad \frac{k_{\rm B}T}{\hbar\omega_{\rm ho}} = 36.93 \; , \quad
 \frac a{d_{\rm ho}} = 0.1215 \; ,
 \label{parameters}
\end{equation}
and keep the hopping constant $J$ fixed. Thus all energies can be scaled with $J$.

\begin{figure}
\centering
\includegraphics{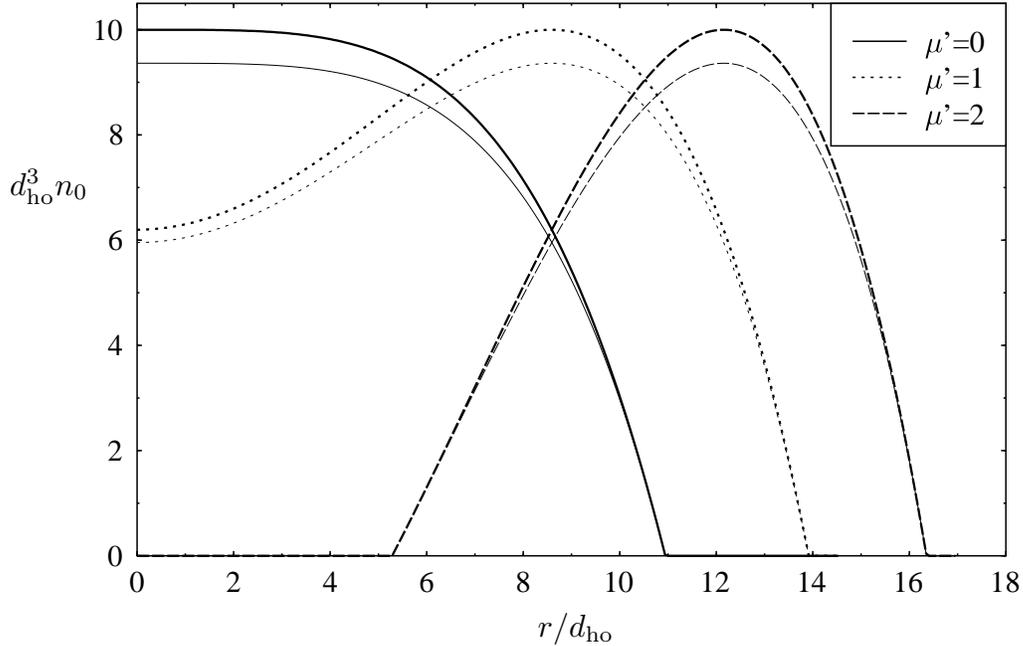}
\caption{Condensate density $n_0=|\Phi|^2/a^3(1+1/\alpha)^2$ of a vortex-free condensate in a 
spherical trap with the numerical parameters given in Eq. (\ref{parameters}) and 
different values of the chemical potential $\mu'$ calculated from the full slave-boson 
mean-field Eq. (\ref{eqTF}) (thick lines) and within RGP approximation (\ref{PhiGP}) (thin lines).}
\label{fig1}
\end{figure}

To calculate the profile of the condensate density in a BEC without vortex, we solve the
TF equation
\begin{equation}
 (1+1/\alpha) - (1+1/\alpha)^2 \frac\partial{\partial|\Phi({\bf r})|^2} \log Z({\bf r})=0 \; .
 \label{eqTF}
\end{equation}
In the RGP approximation (\ref{rgpe}), the solution is
\begin{equation}
 |\Phi_{\rm R}({\bf r})|^2 = \frac{\mu_{\rm R}}{g_{\rm R}} =
 - \frac{a_2(\mu')}{(1+1/\alpha)^2 a_4(\mu')} \; .
 \label{PhiGP}
\end{equation}
Solutions for typical values of the parameters are plotted in 
Fig. \ref{fig1}. The results we get from the renormalised GP 
approximation shows only small deviations
from the numerical solutions of Eq. (\ref{eqTF}).
We find a condensate depletion at the trap center for $\mu'=1$.
This is due to the fact that the condensate is partly suppressed by strong 
interaction effects \cite{kleinert2,dubois1,ziegler2}. For $\mu'=2$ the 
condensate is completely destroyed at the trap center.
We find particle numbers in the condensate of the order
$N_0\approx 10^4\ldots 10^5$.

We note that the total particle density $ n_{\rm tot}$
is much larger than the condensate density $n_0$ and takes values of about 
$0.5 \,a^{-3}$ at the trap center. Thus
the interaction between the non-condensed and the condensed part of the Bose gas plays
a significant role. This implies that the conventional GP Equation, which neglects the
non-condensed part, is not reliable in this parameter regime.

\subsection{Rotating condensate with a single vortex}

In the case of a trap rotating about the $z$-axis with an angular velocity $\Omega$, 
one must include the additional angular momentum
term $-\Omega L_z \Phi({\bf r})$ to the left hand side of the differential equation 
(\ref{SBoriginal}), where $L_z$ is the $z$-component of the angular momentum operator.
This term must also be kept in the
TF approximation. The condensate wave function may develop a vortex then.
We assume here a straight single vortex along the $z$-axis. This
can be described by using cylindrical coordinates and the ansatz
$\Phi({\bf r})=\phi(r_\perp,z) e^{i\varphi}$, where $r_\perp$ is the distance from the
$z$-axis and $\varphi$ the polar angle. The angular momentum operator 
is given as $L_z=-i\frac\partial{\partial\varphi}$. 
This gives rise to an additional term 
$(\Omega/\alpha J-a^2/\alpha(6r_\perp^2))|\Phi({\bf r})|^2/(1+1/\alpha)^2$ in the action
(\ref{Scontinuous}). Instead of Eq. (\ref{eqTF}) we have to solve
\begin{equation}
 (1+1/\alpha) + \left( \frac {a^2}{\alpha 6r_\perp^2} - \frac{\Omega}{\alpha J} \right)
 - (1+1/\alpha)^2 \frac\partial{\partial|\Phi({\bf r})|^2} \log Z({\bf r})=0 \; ,
 \label{eqTFvort}
\end{equation}
and the solution in the RGP approximation is
\begin{equation}
 |\Phi_{\rm R}({\bf r})|^2 = - \frac 1{(1+1/\alpha)^4 a_4(\mu')}
 \left[ \left(\frac{a^2}{\alpha 6r_\perp^2}-\frac{\Omega}{\alpha J}\right)
 + (1+1/\alpha)^2 a_2(\mu') \right] \; .
 \label{PhiGPVort}
\end{equation}

\begin{figure}
\centering
\includegraphics{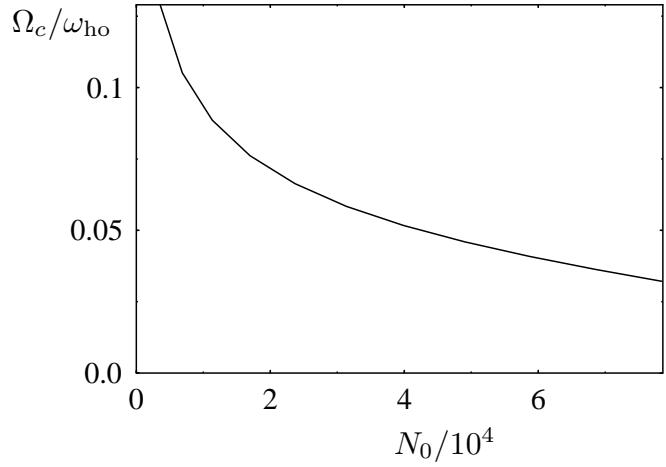}
\caption{Critical angular velocity plotted against the total number of
particles in the condensate.}
\label{OmegaCrit}
\end{figure}

A condensate that is rotating with given angular velocity $\Omega$ forms a stable vortex,
if its total energy is lower than that of a vortex free condensate. This is equivalent
to the condition
\begin{equation}
 S^{\rm vort}(\Omega) - S < 0
\end{equation}
which can be checked numerically by using the TF approximation of Eq. (\ref{Scontinuous})
for a condensate with vortex $S_{\rm FT}^{\rm vort}(\Omega)$ and without vortex $S_{\rm TF}$.
The critical angular velocity $\Omega_{\rm c}$ above which the
vortex is stable is plotted against the number of condensed bosons $N_0$
in Fig. \ref{OmegaCrit}, where $N_0$ is given as
\begin{equation}
 N_0 = \int \frac 1{a^3} |\Phi_{\rm R}({\bf r})|^2 \, {\rm d}^3{\bf r} \; .
\end{equation}
The RGP approximation is in good agreement with the results of the full mean-field
equation.
The decreasing critical angular velocity for higher values of $N_0$ indicates that
a high interaction energy favours the formation of a vortex. It agrees with results
derived from the GP Eq. by perturbation theory \cite{Fetter99} as
well as numerically \cite{Stringari96}.

\begin{figure}
\centering
\includegraphics{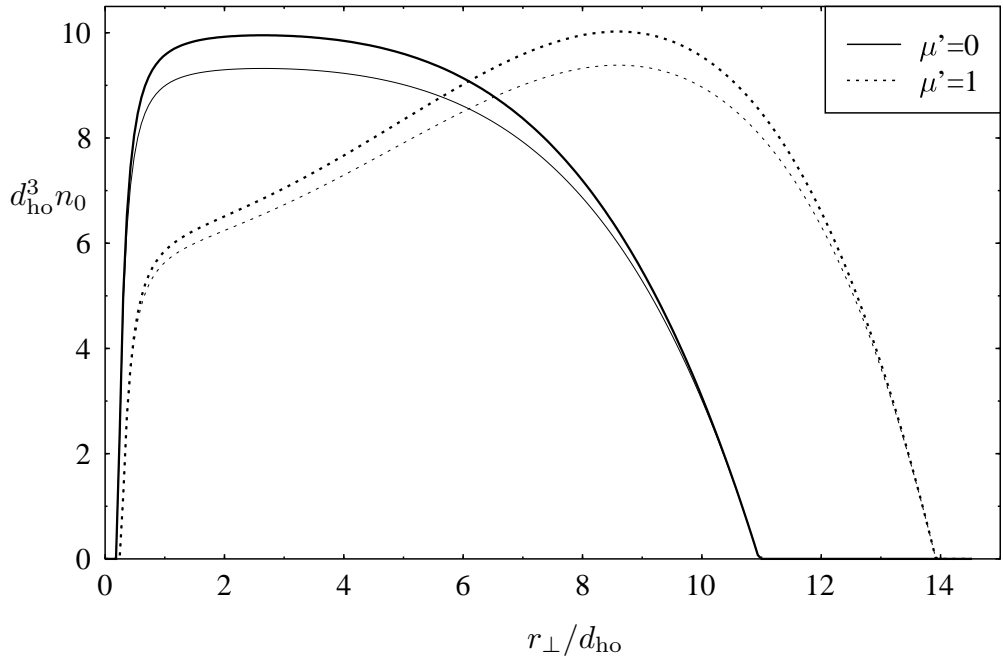}
\caption{Condensate density of a condensate with a single vortex, with same parameters 
as in Fig. \ref{fig1}, calculated from the full mean-field
Eq. (\ref{eqTFvort}) (thick lines) and within RGP approximation (\ref{PhiGPVort})
(thin lines). The rotating frequencies of the trap were chosen to be close to the critical
angular frequency $\Omega_{\rm c}$.}
\label{fig2}
\end{figure}

Typical solutions for shapes of condensate density profiles of BECs with a stable single 
vortex are shown in Fig. \ref{fig2}. In contrast to the case without a vortex,
the condensate is always completely destroyed at the trap center, a feature that also
shows up in the conventional GP approximation \cite{Dalfovo99}.
Again, the RGP approximation is in good agreement with the numerical results from
(\ref{eqTFvort}).

\section{Conclusion}

The slave-boson approach allowed us to study the condensation of a trapped 
high-density Bose gas in a regime where the conventional Gross-Pitaevskii approach is
not valid. Starting from the saddle-point approximation, we have derived a
renormalized Gross-Pitaevskii Eq. with a space dependent coupling constant. This
provides good results in comparison with the more complicated
slave-boson saddle-point calculations. At high
densities, we have found a depletion of the condensate at the trap center due to the
interaction between the condensate and the non-condensate part of the Bose gas. This
feature is not covered by the conventional Gross-Pitaevskii Eq. The
behaviour of the critical angular velocity for the formation of a single vortex agrees
qualitatively with previous results in the literature but supports also the formation of
a vortex for increasing $N_0$.

\section*{Acknowledgements}
The authors want to thank O. Fialko, A. Gammal and K.K. Rajagopal for useful discussions.

\section*{Appendix A}
For a Bose system with creation operators $a_x^+$ and annihilation operators $a_x$ at a
lattice site $x$, an appropriate definition of the condensate density is \cite{leggett}
\begin{equation}
 n_0 := \lim_{x-x'\rightarrow\infty} \left\langle a_x^+ a_{x'} \right\rangle \; .
\end{equation}
In our slave-boson representation, a creation process of a particle is associated to the 
product $b_x^\ast e_x$ and an annihilation process to $e_x^\ast b_x$, thus
\begin{equation}
 n_0 = \lim_{x-x'\rightarrow\infty}
 \left\langle b^\ast_x e_x e_{x'}^\ast b_{x'} \right\rangle \; .
\end{equation}
Here, the expectation value is defined with respect the functional integral (\ref{ZHST}) by
\begin{displaymath}
 \langle\cdots\rangle = \frac 1Z \int\cdots\exp[\ldots]\prod_x
 \delta(|b_x|^2+|e_x|^2-1) {\rm d}b_x {\rm d}b_x^\ast
 {\rm d}e_x {\rm d}e_x^\ast {\rm d}\Phi_x {\rm d}\Phi_x^\ast {\rm d}\varphi_x \; .
\end{displaymath}
We are interested in the connection between the correlation function
$\left\langle\Phi_x \Phi_{x'}^\ast\right\rangle$ and the condensate density.
For this purpose we integrate out the field $\Phi$ to transform the correlation
function of the field $\Phi$ back to a correlation function of the fields $b$ and $e$.
Therefore, we 
perform the integration
\begin{displaymath}
 \fl \beta'^2 \int \Phi_y \Phi_{y'}^\ast \exp\left[ \beta'\sum_{x,x'} \Phi_x ({\bf 1}-\hat{t}')^{-1}
 \Phi_{x'}^\ast + \beta'\sum_x \Phi_x b_x^\ast e_x + \beta'\sum_x \Phi_x^\ast e_x^\ast b_x \right]
 \prod_x {\rm d}\Phi_x {\rm d}\Phi_x^\ast =
\end{displaymath}
\begin{displaymath}
 \fl \frac\partial{\partial(b_y^\ast e_y)} \, \frac\partial{\partial(b_{y'} e_{y'}^\ast)}
 \int \exp\left[ \beta'\sum_{x,x'} \Phi_x ({\bf 1}-\hat{t}')^{-1} \Phi_{x'}^\ast
 + \beta'\sum_x \Phi_x b_x^\ast e_x + \beta'\sum_x \Phi_x^\ast e_x^\ast b_x \right]
 \prod_x {\rm d}\Phi_x {\rm d}\Phi_x^\ast =
\end{displaymath}
\begin{displaymath}
 \fl \frac\partial{\partial(b_y^\ast e_y)} \, \frac\partial{\partial(b_{y'} e_{y'}^\ast)}
 \det \beta'({\bf 1}-\hat{t}') \, \exp\left[
 \beta'\sum_{x,x'} b_x^\ast e_x ({\bf 1}-\hat{t}') e_{x'}^\ast b_{x'} \right] =
\end{displaymath}
\begin{displaymath}
 \fl \beta'^2 \det ({\bf 1}-\hat{t}') \left[({\bf 1}-\hat{t}')_{y,y'} +
 \beta'\sum_{x,x'} b_x^\ast e_x e_{x'}^\ast b_{x'}
 ({\bf 1}-\hat{t}')_{y',x'} \right]
 \exp\left[ \beta'\sum_{x,x'} b_x^\ast e_x ({\bf 1}-\hat{t}')_{x,x'} e_{x'}^\ast b_{x'} \right]
\end{displaymath}
Since we are interested in the limit $y-y'\rightarrow\infty$, and the matrix $\hat{t}'_{y,y'}$
includes nearest-neighbour hopping only, the term $({\bf 1}-\hat{t}')_{y,y'}$ does not
contribute. This yields for far distant lattice points $y,y'$ the expression
\begin{displaymath}
 \left\langle\Phi_y \Phi_{y'}^\ast\right\rangle = \sum_{x,x'} \left\langle b_x^\ast e_x
 e_{x'}^\ast b_{x'} \right\rangle({\bf 1}-\hat{t}')_{x,y} ({\bf 1}-\hat{t}')_{y',x'} \; .
\end{displaymath}
In this sum, only those terms contribute, where $x,y$ as well as $x',y'$ are nearest neighbours.
In the limit $y-y'\rightarrow\infty$ we can assume 
$\left\langle b_x^\ast e_x e_{x'}^\ast b_{x'} \right\rangle=
\left\langle b_y^\ast e_y e_{y'}^\ast b_{y'} \right\rangle$. Thus we can use
\begin{equation}
 \sum_x ({\bf 1}-\hat{t}')_{x,y} = \sum_{x'} ({\bf 1}-\hat{t}')_{y',x'} = (1+1/\alpha)^2 \; .
\end{equation}
We get
\begin{equation}
 \lim_{y-y'\rightarrow\infty} \left\langle \Phi_y\Phi_{y'}^\ast \right\rangle \approx
 (1+1/\alpha)^2 \lim_{y-y'\rightarrow\infty}
 \left\langle b_y^\ast e_y e_{y'}^\ast b_{y'} \right\rangle
\end{equation}
and therefore
\begin{equation}
 n_0 \approx \frac 1{(1+1/\alpha)^2} \lim_{y-y'\rightarrow\infty}
 \left\langle \Phi_y\Phi_{y'}^\ast \right\rangle \; .
\end{equation}
On the mean-field level, this justifies our identification of the condensate denity in 
Eq. (\ref{n0-ident}).

\end{document}